\documentclass[sigconf, screen]{acmart}
\AtBeginDocument{%
  \providecommand\BibTeX{{%
    \normalfont B\kern-0.5em{\scshape i\kern-0.25em b}\kern-0.8em\TeX}}}

\setcopyright{acmcopyright}
\copyrightyear{2022}
\acmYear{2022}
\setcopyright{acmlicensed}\acmConference[CHI '22 Extended Abstracts]{CHI Conference on Human Factors in Computing Systems Extended Abstracts}{April 29-May 5, 2022}{New Orleans, LA, USA}
\acmBooktitle{CHI Conference on Human Factors in Computing Systems Extended Abstracts (CHI '22 Extended Abstracts), April 29-May 5, 2022, New Orleans, LA, USA}
\acmPrice{15.00}
\acmDOI{10.1145/3491101.3519653}
\acmISBN{978-1-4503-9156-6/22/04}

\usepackage{caption}
\usepackage{enumitem}
\usepackage{tabu}
\usepackage{rotating}
\usepackage{tikz}
\usepackage{capt-of,etoolbox}
\usepackage{microtype}
\usepackage{bm}
\usepackage{soul}
\usepackage{wrapfig}
\usepackage{graphbox}
\usepackage{tcolorbox}
\usepackage{cleveref}
\usepackage{mathtools}
\usepackage{balance}
\usepackage[varqu]{zi4}
\usepackage[bb=boondox]{mathalfa}
\usepackage[utf8x]{inputenc}

\usepackage{amsmath}
\usepackage{eqnarray}
\usepackage{bbm}

\definecolor{orange}{RGB}{250,130,49}
\definecolor{red}{RGB}{234,59,90}
\definecolor{agreen}{RGB}{74, 198, 148}
\definecolor{purple}{RGB}{158, 62, 177}
\definecolor{darkpurple}{RGB}{170, 70, 210}
\definecolor{aqua}{RGB}{87, 180, 181}
\definecolor{lightblue}{RGB}{72, 123, 232}
\definecolor{hotpink}{RGB}{255, 83, 115}
\definecolor{teal}{RGB}{90, 200, 250}
\definecolor{linkColor}{RGB}{0, 128, 229}
\definecolor{lightgreen}{RGB}{33, 222, 128}
\definecolor{gray}{RGB}{75, 101, 132}

\definecolor{myred}{RGB}{224, 49, 119}
\definecolor{myorange}{RGB}{250, 130, 49}
\definecolor{myyellow}{RGB}{254, 211, 48}
\definecolor{mygreen}{RGB}{14, 152, 136}
\definecolor{myblue}{RGB}{0, 128, 229}
\definecolor{myviolet}{RGB}{56, 103, 214}
\definecolor{mypurple}{RGB}{136, 84, 208}
\definecolor{mybrown}{RGB}{132, 99, 88}

\definecolor{refColor}{RGB}{109, 35, 130}

\newcommand{\link}[1]{{\href{#1}{\color{linkColor}\textbf{\texttt{#1}}}}}
\newcommand{\figpart}[1]{\textcolor{refColor}{#1}}

\crefname{figure}{fig.}{fig.}
\Crefname{figure}{Fig.}{Fig.}
\crefname{equation}{eq.}{eq.}
\Crefname{equation}{Eq.}{Eq.}
\crefname{section}{\S}{\S}
\creflabelformat{equation}{#2\textup{#1}#3}

\newcommand{\tool}{\textsc{\textsf{StickyLand}}}
\newtoggle{inheader}
\newcommand{\code}{\iftoggle{inheader}{Sticky Code}{\textit{Sticky Code}}}
\newcommand{\codes}{\iftoggle{inheader}{Sticky Codes}{\textit{Sticky Codes}}}
\newcommand{\md}{\iftoggle{inheader}{Sticky Markdown}{\textit{Sticky Markdown}}}
\newcommand{\mds}{\iftoggle{inheader}{Sticky Markdowns}{\textit{Sticky Markdowns}}}

\newcommand{\cell}{\iftoggle{inheader}{Sticky Cell}{\textit{Sticky Cell}}}

\newcommand{\dock}{\iftoggle{inheader}{Sticky Dock}{\textit{Sticky Dock}}}
\newcommand{\fc}{\iftoggle{inheader}{Floating Cell}{\textit{Floating Cell}}}
\newcommand{\fcs}{\iftoggle{inheader}{Floating Cells}{\textit{Floating Cells}}}

\definecolor{soulblue}{RGB}{181, 206, 227}
\definecolor{soulorange}{RGB}{255, 212, 153}
\colorlet{soulmyblue}{myblue!30}

\newcommand{\inlinefig}[2]{\protect\includegraphics[align=c, height=#1pt]{figures/#2.pdf}}

\definecolor{tagbordercolor}{rgb}{0.8, 0.8, 0.8}
\definecolor{tagbgcolor}{rgb}{0.9, 0.9, 0.9}

\newtcbox{\tagg}{nobeforeafter, colframe=tagbordercolor,
colback=tagbgcolor, boxrule=0.5pt, arc=1pt,
  boxsep=0pt,left=2pt,right=2pt,top=1.5pt,bottom=2pt,tcbox raise base}

\begin{document}

\title{\tool{}: Breaking the Linear Presentation of Computational Notebooks}

\author{Zijie J. Wang}
\affiliation{%
  \institution{Georgia Tech}
  \country{}}
\email{jayw@gatech.edu}
\orcid{0000-0003-4360-1423}

\author{Katie Dai}
\affiliation{%
  \institution{Georgia Tech}
  \country{}}
\email{kdai7@gatech.edu}
\orcid{0000-0002-4623-3980}

\author{W. Keith Edwards}
\affiliation{%
  \institution{Georgia Tech}
  \country{}}
\email{keith@cc.gatech.edu}
\orcid{0000-0002-5209-7380}

\begin{abstract}
  How can we better organize code in computational notebooks?
  Notebooks have become a popular tool among data scientists, as
  they seamlessly weave text and code together, supporting users to rapidly iterate and document code experiments.
  However, it is often challenging to organize code in notebooks, partially because there is a mismatch between the linear presentation of code and the non-linear process of exploratory data analysis.
  We present \tool{}, a notebook extension for empowering users to freely organize their code in non-linear ways.
  With sticky cells that are always shown on the screen, users can quickly access their notes, instantly observe experiment results, and easily build interactive dashboards that support complex visual analytics.
  Case studies highlight how our tool can enhance notebook users's productivity and identify opportunities for future notebook designs.
  \tool{} is available at \link{https://github.com/xiaohk/stickyland}.
\end{abstract}



\setlength{\belowcaptionskip}{-12pt}
\setlength{\abovecaptionskip}{5pt}
\begin{teaserfigure}
  \centering
  \includegraphics[width=0.92\textwidth]{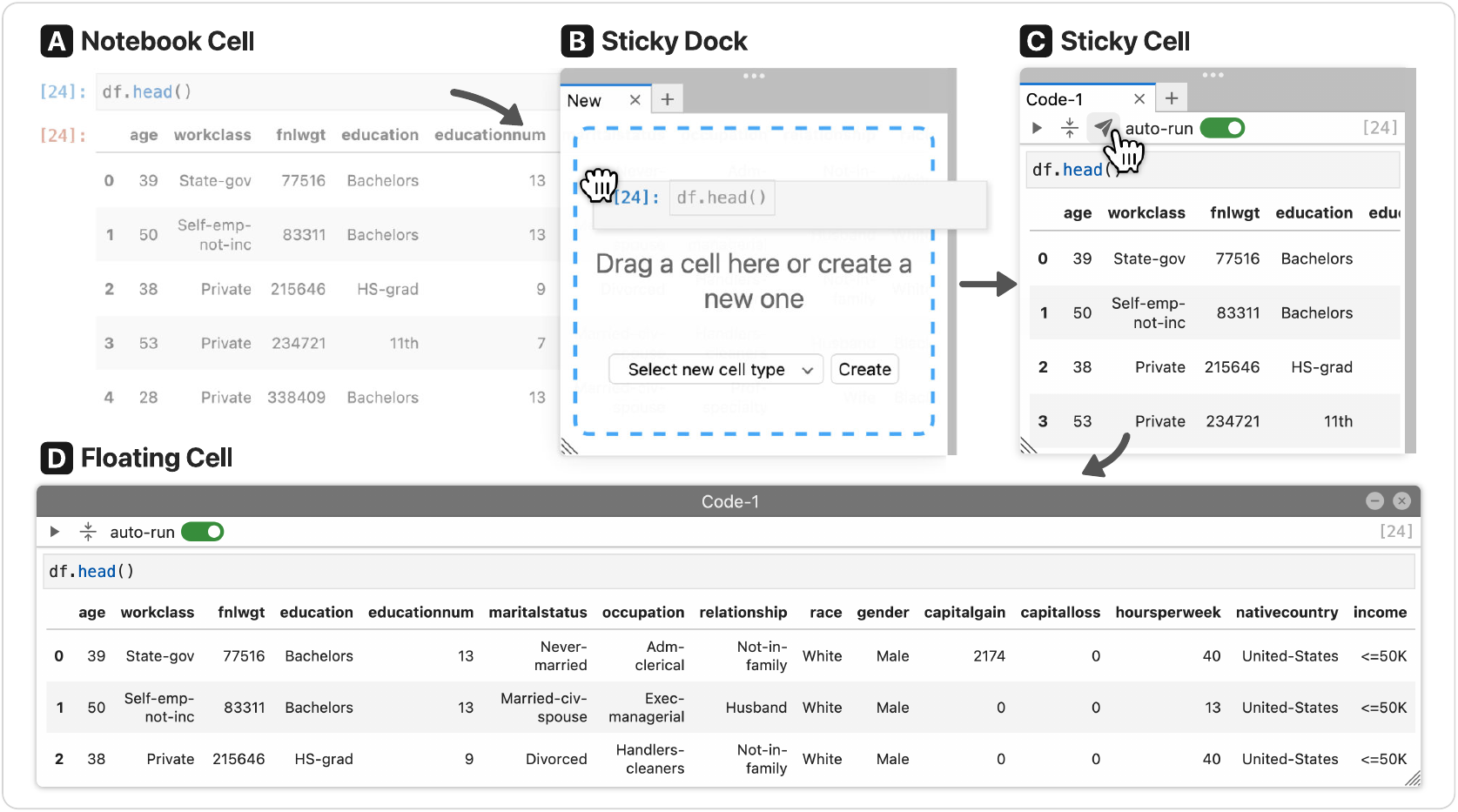}
  \vspace{-5pt}
  \caption{
    The \tool{} user interface persists on top of a computational notebook, creating an always-on workspace for data scientists to display text and execute code.
    (A) A user can drag an existing \textit{Notebook Cell} to (B) the \dock{} that mounts on the edge of the notebook.
    (C) It creates a \cell{} that stays at the same location even when the user scrolls the notebook.
    (D) The user can further convert this cell to a \fc{} that floats on top of the notebook with customizable position and size.
    \looseness=-1
  }
  \vspace{5pt}
  \label{fig:teaser}
\end{teaserfigure}
\setlength{\belowcaptionskip}{0pt}
\setlength{\abovecaptionskip}{12pt}

\maketitle
\section{Introduction}

Computational notebooks, such as Jupyter Notebook~\cite{kluyverJupyterNotebooksPublishing2016} and Colab, have become dominant programming environments for data scientists to explore and understand data~\cite{randlesUsingJupyterNotebook2017}.
According to a 2021 Kaggle survey~\cite{kaggleStateMachineLearning2021}, Jupyter Notebook and JupyterLab are among the top-3 most popular integrated development environments, and more than 75\% surveyed data scientists use Jupyter Notebook or JupyterLab in their day-to-day work.
These notebooks present a customizable programming environment that consists of an arbitrary number of \textit{cells}---small editors for markdown text and code.
These cells are linearly organized in a document format and tend to grow longer and longer as the complexity of the analysis increases.
Textual and visual results, such as code output and visualizations, are presented below corresponding cells.
Users can arrange and execute these cells in their notebooks---creating a form of literate programming~\cite{knuthLiterateProgramming1984}.
With a seamless combination of text, code, and visual outputs, notebooks facilitates data scientists to \textit{perform} exploratory data analysis, \textit{document} data insights, and eventually \textit{share} data stories with collaborators.

However, recent research suggests that notebook users face challenges regarding code organization~\cite{chattopadhyayWhatWrongComputational2020,headManagingMessesComputational2019}.
Since there is only one code interpreter that keeps all the execution states in a notebook, users need to carefully arrange and execute cells in certain orders to avoid variable corruptions~\cite{headManagingMessesComputational2019, keryEffectiveForagingData2019}.
In addition, all notebooks follow a linear presentation style that is contradictory to the non-linear nature of exploratory data analysis~\cite{keryStoryNotebookExploratory2018,weinmanForkItSupporting2021}.
Effectively, all of the code cells are a part of a single program, despite being spread across the notebook.
This means that edits such as changing variables in one part of the notebook may have unforeseen consequences later in the notebook.
To help users better organize notebook cells, researchers propose techniques such as distilling essential cells to create sub-notebooks~\cite{headManagingMessesComputational2019}, supporting cell-level code version controls~\cite{mikamiMicroVersioningToolSupport2017}, and two-column notebook layout~\cite{weinmanForkItSupporting2021}.
In contrast to these works, we address the notebook messiness by focusing on the linearity of the traditional notebook layout.

We introduce \tool{}~(\autoref{fig:teaser}), an interface that breaks the linearity of notebooks with \textit{sticky cells} that are always shown on top of the notebook.
This alternative layout provides multiple new options for \textit{organizing}, \textit{navigating}, \textit{displaying} and \textit{executing} notebooks.
Our main contributions are:

\begin{itemize}[topsep=5pt, itemsep=3pt, parsep=0mm, leftmargin=4mm]
    \item \textbf{\tool{}, a collection of user interface techniques} that empowers data scientists to break the linear presentation of traditional notebooks.
    These user interface techniques include the ability to create persistent, free-floating cells that can contain either code or markup, the ability to automatically run code cells in order to help synchronize states and explore results instantly; and a variety of organizational techniques that let users rearrange and combine sticky cells in order to create customized dashboards that support complex visual analytics.
    We also present three use scenarios where \tool{} enhances notebook users' productivity, collaboration, and learning through more flexible code organizations.

    \item \textbf{An open-source\footnote{\tool{} Code: \link{https://github.com/xiaohk/stickyland}} implementation} that broadens the public's access to a more flexible notebook layout.
    We also provide comprehensive documentation to help future designers and researchers use \tool{} as a user interface toolkit to explore and develop alternative notebook designs.
    To see a demo of \tool{}, visit \link{https://youtu.be/OKaPmEBzEX0}.
\end{itemize}

\noindent We hope our work will inspire the design, research, and development of computational notebooks that help data scientists to more productively analyze data, document insights, and share findings.

\section{Background and Related Work}

As computational notebooks have been gaining popularity in recent years, there is a growing body of research from the CHI community that aims to improve these notebooks~\cite[e.g.,][]{keryEffectiveForagingData2019, ruleExplorationExplanationComputational2018,keryStoryNotebookExploratory2018}.
Through conducting interviews and surveys with data scientists, researchers have identified difficulties of using notebooks in practice: One of the main pain points is code management~\cite{chattopadhyayWhatWrongComputational2020, keryStoryNotebookExploratory2018}.
Since the traditional scripting code is modularized as a linear collection of multiple cells that users can arrange and execute in any order, it becomes challenging to manage code in computational notebooks~\cite{chattopadhyayWhatWrongComputational2020, lauDesignSpaceComputational2020}.
For example, the fact that all the code cells are effective in one program means that a change to some code at the top of the notebook may impact or break code far away at the end of the notebook.
Likewise, as the data analysis gets more complex, users may find themselves flipping back and forth through the long notebook, trying to find specific pieces of information.\looseness=-1

To address this issue, researchers introduce techniques to clean unused code~\cite{headManagingMessesComputational2019, ruleExplorationExplanationComputational2018} and provide better version control of code~\cite{keryVarioliteSupportingExploratory2017, mikamiMicroVersioningToolSupport2017}.
To align notebook designs with the non-linear and iterative nature of exploratory data analysis, \citet{weinmanForkItSupporting2021} explore alternatives to the single execution state of notebooks with forking and backtracking, as well as a two-column layout.
Similarly, researchers have also explored designs that allow users to more easily navigate between code and interactive visualizations, such as bi-directional communications~\cite{keryMageFluidMoves2020} and side-by-side presentations~\cite{wuB2BridgingCode2020}.
However, these works still follow the traditional linear presentation style of notebooks.
Such linear presentation is not the only way to organize code; In fact, before computation notebooks, researchers developed a wide range of tools that support visual organization of code, such as partially linked code in a grid~\cite{hartmannDesignExplorationCreating2008a}, stackable cards to dynamically organize code segments~\cite{nelsonIDESupportProgramming2017}, single code editor across multiple documents~\cite{hanTextletsSupportingConstraints2020}, and code bubbles that users can freely group and arrange~\cite{bragdonCodeBubblesWorking2010}.
Inspired by these works, \tool{} challenges the linear layout of notebooks by introducing flexible sticky cells that persist on top of the notebook.
To help users easily inspect code results and navigate the notebook, \tool{} supports automatically running cells and a variety of organization techniques.

\section{System Design and Implementation}

\tool{} is an interface that breaks the traditional linear presentation of computational notebooks by introducing a persistent area on top of the notebook where users can freely store any notebook artifacts such as code, notes, and task lists.
It can help users more easily navigate the notebook, write less repetitive code, and better manage the execution states of their code.
To accomplish this, \tool{} starts with the \dock{}~(\autoref{sec:dock}), a persistent area shown on top of the notebook that contains sticky cells.
With easy drag-and-drop, users can add editable \code{} cells~(\autoref{sec:code}), which can execute code and have an auto-run feature and \md{} cells~(\autoref{sec:md}), which can format markdown and \LaTeX{} text.
In addition, users can create multiple \fcs{}~(\autoref{sec:float})---enabling more flexible code display and easy dashboard creation. The following sections describe these features in detail.

\begin{figure}[tb]
  \includegraphics[width=\linewidth]{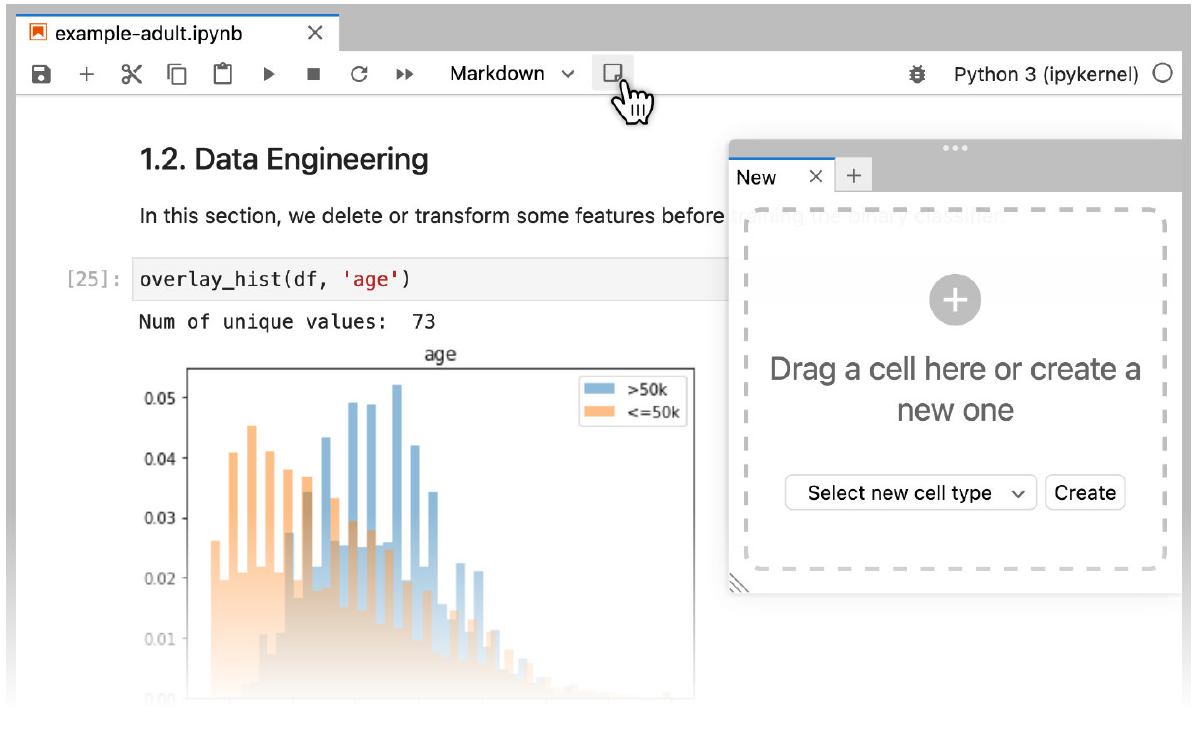}
  \Description{Auto-run feature image}
  \caption{
    The \dock{} sticks on a notebook's right edge, containing multiple sticky cells as tabs.
    To add new cells, users can drag-and-drop existing cells from the notebook or directly create new sticky cells from scratch.
  }
  \label{fig:dock}
\end{figure}

\subsection{Sticky Dock}
\label{sec:dock}

When users launch \tool{} in a notebook by clicking a button in the toolbar, the \dock{} appears---sticking to the right edge of the notebook window~(\autoref{fig:dock}).
Users can resize the dock by dragging the panel corner and re-position it by dragging the top handle up and down along the right window edge.
\dock{} is always shown on top of the main notebook, and it does not move when users scroll the notebook.
\dock{} consists of a tab bar where each tab is associated with one sticky cell.
Users can create an unlimited number of tabs to store sticky cells.
To create a new tab, users can click the \texttt{+} button which opens a new tab with a dropzone.
Users can easily drag-and-drop existing cells from the main notebook to the dropzone.
If an existing cell is dragged to \dock{}, \tool{} creates a clone of the cell and \textit{collapses} (hides) the original cell in the notebook.
\tool{} synchronizes the content (both input and output results) in the existing cell and the cloned cell.
In other words, editing a cell also changes the original cell in the notebook in real-time.
Users can also create a new sticky cell from scratch in the dropzone, and \dock{} will add a collapsed cell in the notebook to synchronize the content with the new sticky cell.
Once users close a tab, \dock{} removes the sticky cell and \textit{extends} (shows) its corresponding cell in the notebook.

\subsection{Sticky Code}
\label{sec:code}

\begin{figure*}[tb]
  \includegraphics[width=\linewidth]{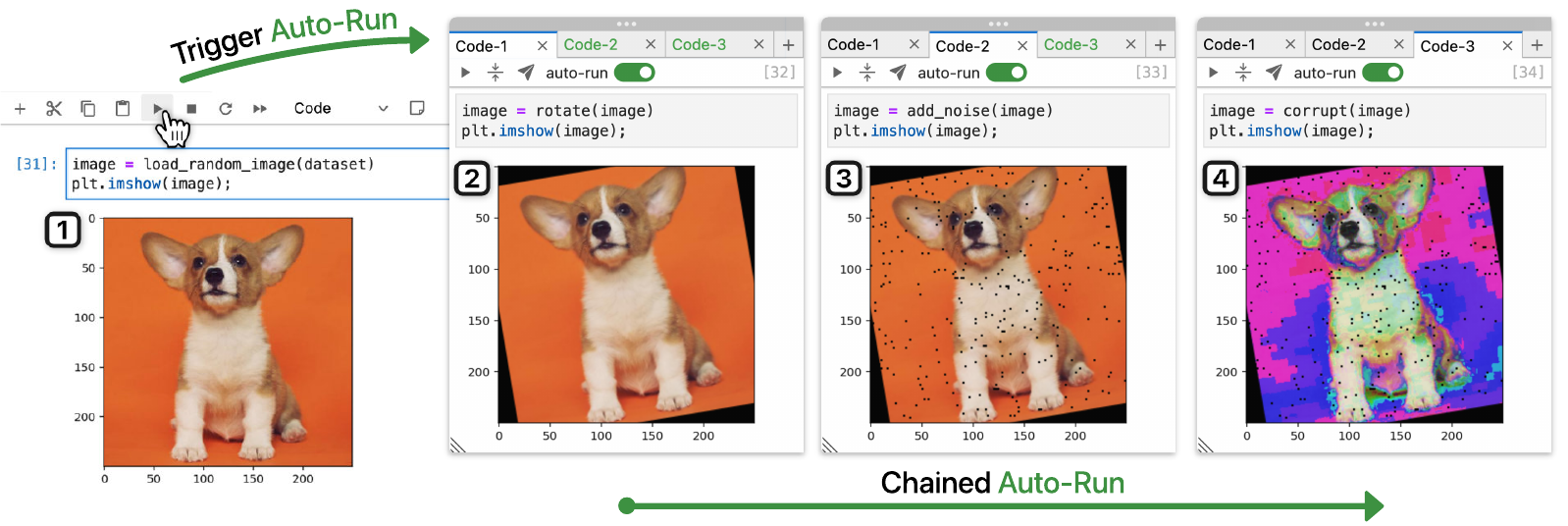}
  \Description{Auto-run feature image}
  \caption{
    To help users avoid repetitively running the same code, the \code{} supports to automatically execute its code when any other code cell is executed.
    Take a data scientist who inspects their image augmentation pipeline for example;
    \inlinefig{10}{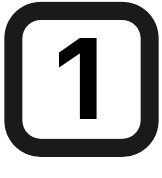} They first run a code cell in the main notebook that loads a random image from their dataset.
    This interaction triggers three \codes{} with \texttt{auto-run} toggled to automatically execute in a \textit{chained} manner: \inlinefig{10
    }{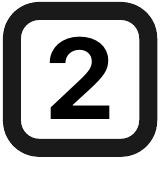} first rotating the original image; \inlinefig{10}{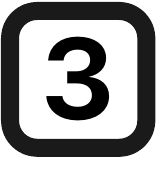} then adding noise to the rotated image; \inlinefig{10}{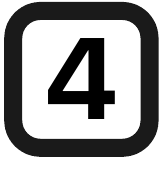} finally corrupting the edited image.
  }
  \label{fig:autorun}
\end{figure*}

In the \dock{}, users can create a tab that contains a traditional code cell, where each cell has an input and output cell~(\autoref{fig:autorun}).
In computational notebooks, users write scripts (e.g., Python, R, Julia) in the input cell, and the output cell displays the result after users execute the input cell.
The output cell can render many different forms of data, from simple strings, tables, and charts to any HTML markups.
Therefore, users can use the \code{} to display online videos and interactive visual analytics tools~\cite{wexlerWhatIfToolInteractive2019, tenneyLanguageInterpretabilityTool2020, liArgoLiteOpenSource2020}.

Users can easily create new \code{} cells by using drag-and-drop to copy cell code from the notebook to \tool{}.
Users can also create a new \code{} from scratch.
The \code{} has a cell synchronization feature: code in the \code{} inherits and shares the code and execution states from the notebook, so editing and executing the cell in \tool{} will make the same changes to the content and execution states in the notebook as well.
In other words, users can access variables defined in the notebook from \tool{}, and vice versa.
\code{} shares the global code state so that it is easier for users to keep a mental model of the code execution history across the notebook.

On top of the input cell in the \code{}, there is a toolbar with four buttons, where users can click to run, hide the input cell, make the cell float~(\autoref{sec:float}), and toggle the \textit{auto-run} mode.
When the auto-run mode is active, every time the user runs a different cell from the notebook or \tool{}~(\autoref{fig:autorun}\figpart{-1}), the \code{} automatically runs the input cell and displays the output~(\autoref{fig:autorun}\figpart{-2}).
The auto-run feature helps users avoid manually running cells that need recurring updates, such as code that displays the results of code experiments.
If there are multiple \codes{} in the auto-run mode, \tool{} will automatically run them in the order they are created~(\autoref{fig:autorun}\figpart{-2$\sim$4}).
\dock{} also highlights auto-run \code{} that has unseen updates with a green tab name.

\subsection{Sticky Markdown}
\label{sec:md}

The \md{} is similar to the \code{}~(\autoref{sec:code}): users can type Markdown code in the input cell, and then the output cell will render the corresponding formatted textual result.
Users can use a keyboard shortcut or toolbar buttons to switch between the input and output cells.
Similarly to the \code{}, users can make the \md{} float, and the content of a \md{} is synchronized with its corresponding cell in the notebook.
Users can also render LaTeX mathematical equations in the input cell.
With \mds{}, users can easily store artifacts such as data exploration notes, code snippets, and to-do lists.

\subsection{Floating Cells}
\label{sec:float}

There is a ``launch'' button in the toolbar of both \code{} and \md{}~(\autoref{fig:autorun}\figpart{-2}), where users can click to launch the sticky cell outside of the \dock{} container~(\autoref{fig:teaser}\figpart{D}).
Then, the cell transitions into a separate cell window with a smooth animation.
This \fc{} floats on top of the notebook, and it does not move when users scroll the notebook.
During the transition, the cell window is auto-resized to fit the input cell size.
Users can drag the header of \fc{} to change its position and drag the left corner to resize the window.
Floating \code{} and \md{} keep all the same functionalities including auto-run as they are in the \dock{}.
Users can minimize or close the \fc{} by clicking the buttons on its header.

In addition, \tool{} allows users to launch multiple \fcs{} on the screen.
Therefore, with a combination of floating \code{} and \md{} as well as the auto-run feature, users can easily create an interactive dashboard.
With different components on a dashboard, users can display text and visual outputs, provide input values, run code, or interact with stand-alone interactive tools.
As users can freely customize and arrange \fcs{}, users can use the interactive dashboard for data analysis or as a storyboard to communicate data insights to other people.

\subsection{Open-source Implementation}
\label{sec:implement}

We implement \tool{} as a JupyterLab Extension.\footnote{JupyterLab Extension:\link{https://github.com/jupyterlab/extension-examples}}
With JupyterLab's flexible and extensible design, we can customize the presentation styles of notebooks.
Through extension APIs, \tool{} has access to cell content, and it can listen to code execution events from the notebooks.
With a responsive design and integrated style as native UI elements, we hope \tool{} can provide data scientists a seamless experience when using JupyterLab for exploratory data analysis.
We also release \tool{} as a Python package so that users can easily install it with one command.

\section{Usage Scenario}

\subsection{Exploratory Data Analysis}

\setlength{\columnsep}{8pt}%
\setlength{\intextsep}{-3pt}%
\begin{wrapfigure}{R}{0.24\textwidth}
  \vspace{0pt}
  \centering
  \includegraphics[width=0.24\textwidth]{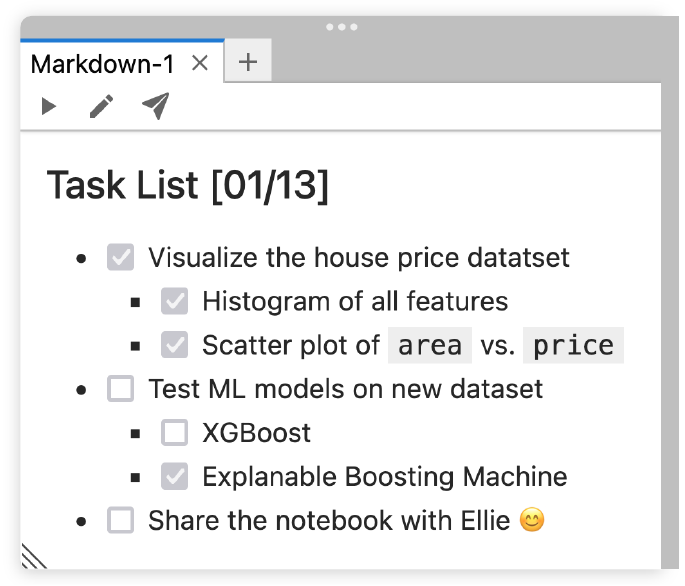}
  \vspace{-10pt}
  \label{fig:md}
\end{wrapfigure}
As a junior data scientist, Carl struggles with organization and keeping track of his thoughts and findings.
With \tool{}, he can keep everything in the same place for easy access.
Carl is working on a complicated task with many steps, so he opens \tool{} and creates a new tab with a new \md{} cell from scratch.
He double clicks the input cell to enter the editing mode and adds his to-do list (shown on the right); it helps him plan his work process.
As he keeps working, he can always see the list of steps that helps him remember and follow his plan.
Carl also clicks the rendered checkbox to check-off completed items and update the list as needed.
As he continues conducting the data analysis, he creates a new \md{} tab and writes down his thoughts and findings.
The next day, he has a ``gotcha'' moment at work, so he quickly opens \tool{} as a reference to remember his findings from the day before and make comparisons.
Here, \tool{} acts like a miniature notebook that is always accessible, as it persists on the screen regardless of the tasks happening in the notebook.
Now Carl wants to test and compare some results to confirm some hypotheses regarding the dataset, but his visualization results are buried in different sections within his long notebook.
To better organize these visualizations, Carl creates several \codes{} tabs and drags related code into \tool{}.
He also collapses the input cell so that only the visualization output is visible.
Carl turns on the auto-run mode on all visualization \codes{}, so that every time he makes a change in the notebook, the visualization refreshes.
Now Carl can focus on writing code in the main document to try out different data transformation methods, as he can always inspect the latest experiment results in \tool{}.
With accessible and automatically-updated visualizations, Carl can quickly verify his hypotheses.

\subsection{Communicating Machine Learning Models}
\label{sec:dashboard}

\setlength{\belowcaptionskip}{0pt}
\setlength{\abovecaptionskip}{5pt}
\begin{figure*}[tb]
  \includegraphics[width=\linewidth]{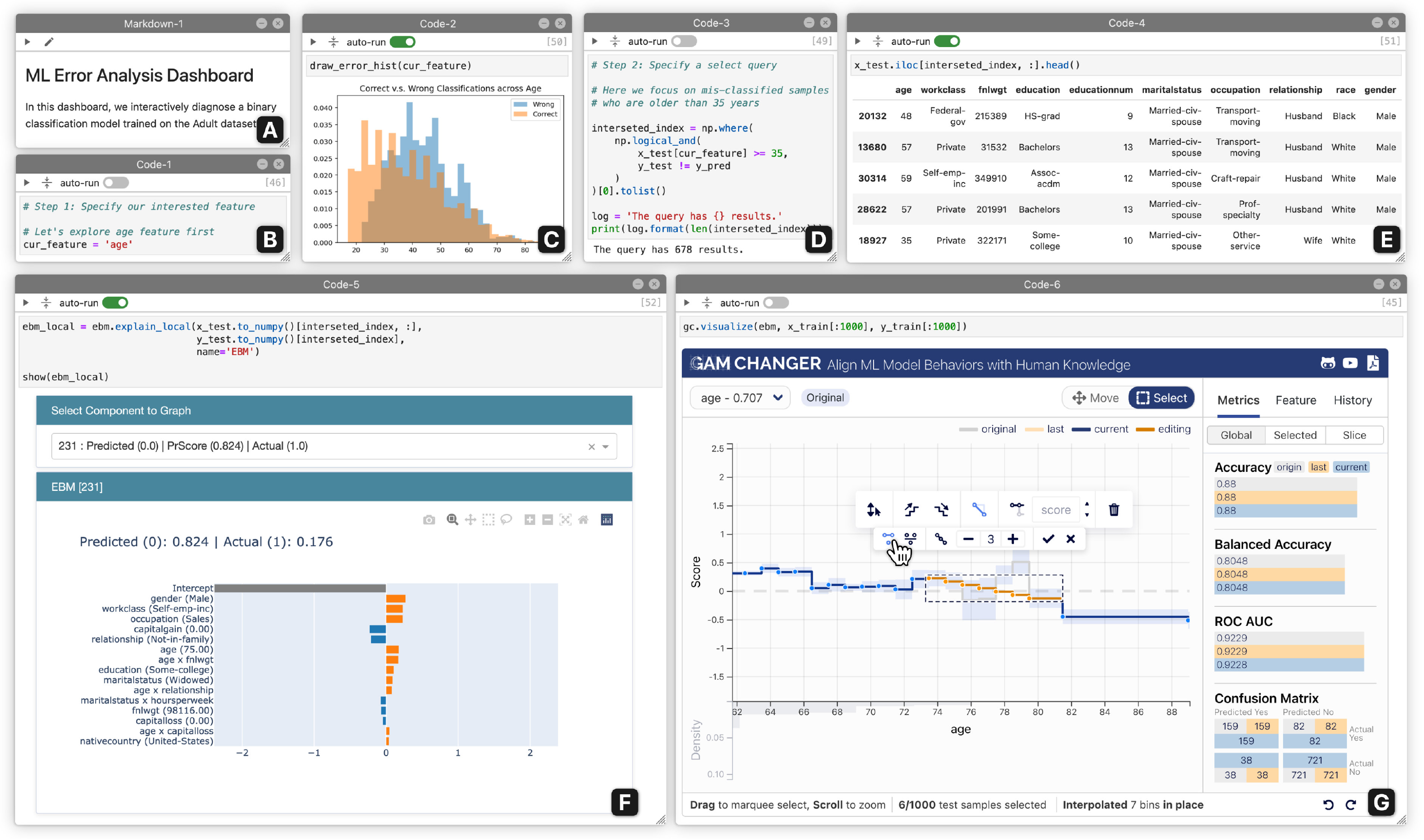}
  \Description{Auto-run feature image}
  \caption{
    With multiple resizable and re-positionable \textit{\fc{}s} that float on top of the notebook, a user can quickly create an interactive dashboard that support complex visual analytics directly from their existing code.
    For example, to diagnose a machine learning (ML) model's performance on different dataset slices, an ML engineer can create a dashboard to interactively perform error analysis through simple drag-and-drops.
    The dashboard consists of: \inlinefig{9}{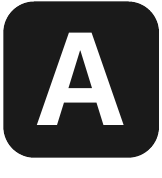} \textit{markdown text} describing the dashboard, \inlinefig{9}{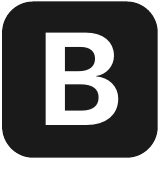} \textit{input field} to specify a feature to diagnose, \inlinefig{9}{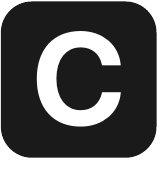} auto-run \textit{chart} showing the distribution of the specified feature, \inlinefig{9}{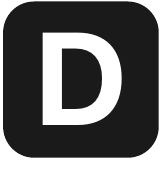} second \textit{input field} to further specify the range within the feature to diagnose, \inlinefig{9}{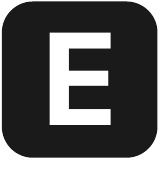} auto-run \textit{table} displaying all samples that meet the criteria, \inlinefig{9}{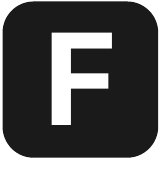} auto-run \textit{visualization} explaining how the ML model makes decision on these samples, \inlinefig{9}{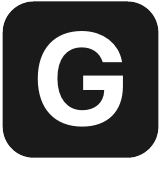} \textit{interactive tool} allowing the ML engineer to fix the ML model by editing its parameters based on their error analysis.
  }
  \label{fig:dashboard}
\end{figure*}
\setlength{\belowcaptionskip}{0pt}
\setlength{\abovecaptionskip}{12pt}

Ellie, a senior machine learning (ML) engineer at a financial institution, has trained an ML model that helps the institution make loan application decisions.
This model takes inputs of an individual's financial information, such as credit score, annual income, and employment length, to inform a lending decision.
As this model could directly affect the livelihood of thousands of clients, Ellie wants to make sure that the model's behavior aligns with the financial experts' knowledge and does not discriminate against specific demographics.
Ellie develops her model in JupyterLab, so she decides to use \tool{} to create an interactive dashboard (\autoref{fig:dashboard}) to perform ML model error analysis with a financial expert.
She first creates a \md{} that describes the purpose of the dashboard, and then she clicks the launching button to convert the cell to a \inlinefig{9}{icon-a} \fc{} (\autoref{fig:dashboard}\figpart{A}) that floats on top of the notebook.
She drags the \fc{} to the top left of the notebook and resizes it accordingly.
Following the same procedure, Ellie further creates:\looseness=-1

\begin{itemize}[topsep=3pt, itemsep=1pt, parsep=0mm, leftmargin=6mm, label={}]
    \item[\inlinefig{9}{icon-b}] \textbf{Code cell} that allows users to specify a feature from the dataset on which to perform error analysi
    \item[\inlinefig{9}{icon-c}] \textbf{Histogram chart} with auto-run that visualizes the distribution of all samples on the specified feature
    \item[\inlinefig{9}{icon-d}] \textbf{Code cell} that allows users to further specify a range with the selected feature to diagnose the model
    \item[\inlinefig{9}{icon-e}] \textbf{Table} with auto-run that displays the details of all samples from the dataset that meet the specified range
    \item[\inlinefig{9}{icon-f}] \textbf{Interactive chart} with auto-run that explains how the model makes the decision on these samples~\cite{noriInterpretMLUnifiedFramework2019}
    \item[\inlinefig{9}{icon-g}] \textbf{ML editor} that helps users align a model's behavior with their knowledge by editing the model's parameters~\cite{wangGAMChangerEditing2021}
    \looseness=-1
\end{itemize}
\noindent Then, Ellie and the financial expert examine the model's behaviors across common under-represented demographics (e.g., older people and people of color) by changing the feature and feature range in two code cells (\autoref{fig:dashboard}\figpart{B, D}).
Once Ellie runs these two cells, all other cells in the dashboard with auto-run toggled automatically refresh themselves and display information in sync with the specified subset of data.
Finally, they use the model editor (\autoref{fig:dashboard}\figpart{G}) to create a more equitable model by editing its parameters to mitigate biases they discover from this dashboard.
With \tool{}, Ellie is now able to quickly build an interactive and tightly-integrated dashboard that supports collaborative and complex visual analytics.\looseness=-1

\subsection{Learning Programming}

Russel, an undergraduate computer science student, is taking a class to learn how to use Python for data analysis.
His instructor requires students to submit their assignments in Jupyter Notebooks.
The starter notebook lists detailed instructions at the top followed by a long list of skeleton code.
When completing the assignments, Russel usually has to scroll all the way back up to check the instructions.

\setlength{\columnsep}{10pt}%
\setlength{\intextsep}{-10pt}%
\begin{wrapfigure}{R}{0.23\textwidth}
  \vspace{-3pt}
  \centering
  \includegraphics[width=0.23\textwidth]{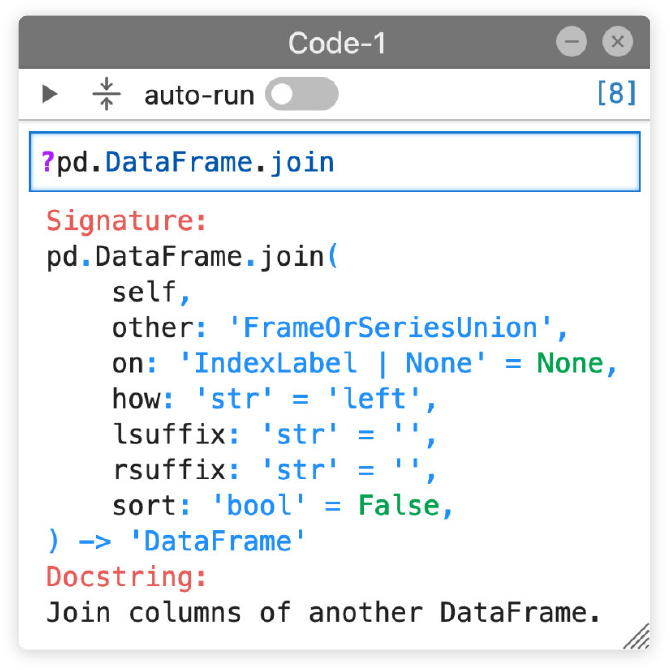}
  \label{fig:md}
\end{wrapfigure}
\noindent After installing \tool{} this week, he realizes he can drag the instructions into a new \md{} to store the steps to refer to as he works through the assignment.
Russel is struggling with completing one problem as he keeps forgetting the syntax and how to use some Python functions.
To get help, Russel creates a new floating \code{} in \tool{}, where he imports a YouTube video of a Python tutorial with code walk through.
He then drags that \code{} cell to resize the video so that he can read the code in the video.
Seeing the video while writing code at the same time helps Russel solve many problems.
Later, Russel gets frustrated that he keeps forgetting the parameter list of some functions.
Therefore, he creates another floating \code{} where he prints out the documentation of confusing functions (shown on the right).
This sticky cell enables Russel to quickly refer to the syntax of functions while completing the assignment.
With \tool{}'s help, Russel quickly finishes this week's assignment and eagers to learn more.

\section{Limitations \& Future Work}

While \tool{} can help users better organize their code in computational notebooks, there are some potential improvements to our current system design regarding the auto-run feature~(\autoref{sec:limitation:auto}), code states~(\autoref{sec:limitation:state}), and notebook platforms~(\autoref{sec:limitation:notebook}).
We also plan to conduct a user study to evaluate \tool{}~(\autoref{sec:limitation:study}).

\subsection{Support Topological Auto-run}
\label{sec:limitation:auto}

We develop the auto-run feature in \code{} to help users avoid running cells that need recurring updates~(\autoref{sec:code}).
When auto-run is activated, a \code{} would automatically execute its code after any other cell is executed.
However, this mode can fail in some special cases.
For example, an auto-run \code{} A depends on cell B which depends on cell C.
If a user runs cell C, then the \code{} A can yield unexpected results after its automatic execution, because cell B is not updated.
Also, an auto-run \code{} might waste unnecessary execution time when a user runs unrelated cells.
To address this issue, we plan to take inspirations from Observable\footnote{Observable: \link{https://observablehq.com}} and support \textit{topological automatic execution}: auto-run \code{} would only run when their referenced values changed in other cells, and it would also automatically recompute all intermediate cells that uses the same referenced values in a topological order.

\subsection{Explore Separate States}
\label{sec:limitation:state}

We are working on adding the feature of separating the code state in \tool{} from the cell state in the main notebook.
This feature would be a unidirectional state sharing mechanism, where the code in the \code{} inherits states from the notebook, but the code in the notebook does not have access to the states in the \code{}.
In other words, instead of creating \code{} that live in one big program as the notebook, these cells would be independent from each other, with their own scope, thus providing better abstraction, freedom from side effects, and separation of concerns.
For example, users can create visualizations in the \code{} using data from the notebook, but they cannot access variables defined in the \code{} from the notebook.
By partially separating the execution states, \tool{} might help avoid information overload when tracking the code states across the notebook.
For example, a user can use cells with separate states as a ``safe space'' for throw-away experiments, as they do not need to worry about interfering with the states in the notebook.
Thus, having separate code states might further challenge the linearity of computational notebooks.

\subsection{Generalize to Other Notebooks}
\label{sec:limitation:notebook}

For future work, we would look at generalizing \tool{} to be used for other computational notebooks such as Google Colab, VSCode Notebook, Azure Notebook.
While Jupyter is the most popular computational notebook, expanding \tool{} to other notebooks with similar functionality would increase benefits to a wider range of users.
\tool{} could then help more users who work in different domains with different programming languages.

\subsection{Planned Evaluation}
\label{sec:limitation:study}

We are preparing to conduct a survey study to evaluate the usability of \tool{}.
We will recruit data scientists who use JupyterLab or Jupyter Notebook in their work through social media and word of mouth.
Then, we will provide participants with information regarding the features of our tool and detailed instructions on how to use it in notebooks with a tutorial video.
We will ask participants to use \tool{} in their daily work.
After some time, we will ask participants to complete a survey to (1) report their daily usage and use patterns, (2) evaluate the usability of our tool, and (3) provide feedback on the design and implementation.

\section{Conclusion}
As computational notebooks have become the most popular programming tool among data scientists, it is critical to address the limitations of the current notebook design.
In this work, we present \tool{}, a user interface that leverages an always-on workspace to break the linear presentation of traditional notebooks.
We implement our tool as an open-source JupyterLab extension.
It enables users to easily install and use \tool{} in their daily work.
Also, future designers and developers can use our tool to explore alternative notebook designs and implementations.
We discussed use scenarios where users with different levels of familiarity with data science can benefit from a non-linear layout of computational notebooks.
We hope our work will inspire future researchers to study and evaluate alternative notebook designs that help improve the usefulness, productivity, and user experience for notebook users and eventually help democratize data science.

\begin{acks}
We thank Haekyu Park, Rahul Duggal, Duen Horng (Polo) Chau, Benjamin Hoover, and Seongmin Lee for the fruitful discussions.
We appreciate the Project Jupyter community for answering our JupyterLab extension development questions.
We are also very grateful to anonymous reviewers for their valuable feedback.
\end{acks} 
\bibliographystyle{ACM-Reference-Format}
\bibliography{notebook}

\end{document}